# +FAULT TOLERANCE IN REAL TIME MULTIPROCESSORS - EMBEDDED SYSTEMS

A. Christy Persya, Sr.Lecturer,
Department of Information Science and Engineering
The Oxford College of Engineering
Bangalore, India
christypersya@gmail.com

T.R.Gopalakrishnan Nair Ph.D.
Director, Research and Industry,
DS Institutions
Bangalore, India
trgnair@ieee.org

## ABSTRACT

All real time tasks which are termed as critical tasks by nature have to complete its execution before its deadline, even in presence of faults. The most popularly used real time task assignment algorithms are First Fit (FF), Best Fit (BF), Bin Packing (BP).The common task scheduling algorithms are Rate Monotonic (RM), Earliest Deadline First (EDF) etc.All the current approaches deal with either fault tolerance or criticality in real time. In this paper we have proposed an integrated approach with a new algorithm, called SASA (Sorting And Sequential Assignment) which maps the real time task assignment with task schedule and fault tolerance.

## 1. INTRODUCTION

An embedded system is a special purpose computer system designed to perform one or a few dedicated functions with real-time computing. Real time computing means that all the task have to finish its execution before its deadline. Examples of such computing applications are systems that control trucks, trains, aircraft, satellites and industrial process control systems. A task is a collection of related jobs (or instances) cooperating to executea function. The tasks in the system can be represented by models. Tasks can be periodic, aperiodic or sporadic tasks. The release time of the tasks is random and cannot be predicted. Tasks can be executed based on some priorities (deadline).Priorities can be either fixed or dynamic. Static scheduling and dynamic scheduling is possible.

Real time systems are characterized by three main features: 1) response time 2) Fault Tolerance 3)Task Scheduling. Response time is the operation correctness of a real-time system which depends not only on its logical results, but also on the time at which these results become available. Fault tolerance, means all the tasks admitted to the system completes its execution even in the presence of faults. Task scheduling, is the process to decide on when and which processor the given tasks should be executed [5].This paper has three main contributions:1)SASA(Sorting And Sequential Assignment) algorithm 2)EDF Scheduling 3) mapping FT-EDF.

The rest of the paper is organized as follows. Section 2 describes the task splitting on multiprocessors with an algorithm SASA.Section 3 discusses the task execution on each multiprocessor along with an algorithm.

In section 4, the mapping of fault tolerance with real time scheduling is discussed and section 5 explains the conclusion of the paper.

## 2. TASK ASSIGNMENT

The SASA algorithm given below can be used for assigning the tasks to the processors. It assumes that the tasks are sorted so that Ti<Ti+1.The concept used in the algorithm is as follows: First all the tasks are been sorted in ascending order. Then in sequential manner, all the sorted tasks are assigned to each processor such that processor utilization does not exceed the threshold value. If any task is not able to be assigned to any of the processors then the task can be split into two and assigned to two different processors as two different tasks .These tasks can be executed such a way that, as soon as first part of the task completes its execution it preempts the next half in the other processor.

| Algorithm for Task Assignment |
|---|
| 1. Sort the Task Set T such that T1<=T2<=…………..<=Tn.<br>2. Initialize the processor utilization upper bound variable UTH to the threshold value.<br>3. Compare the processor Ui Utilization (When Ti is assigned) to the UTH.<br>4. If it is less, then assign the task Ti on Pi.<br>5. Else repeat on Pi+1 from step 3.<br>6. If a task Tm is not able to be assigned on any of the processor, then split the task and assign it to different processor such that Ui of the processor I should not exceed UTH.<br>7. Repeat step 6 for all remaining tasks |

The earlier existing algorithm does not sort the tasks based on the periods before assigning the tasks.Shinpei Kato [1] has given an algorithm for sorting the tasks but not to achieve fault tolerance. This algorithm provides a path to sort the tasks and assign all the tasks to all multiprocessors using Bin Packing algorithm to achieve fault tolerance.

## 3. TASK EXECUTION

EDF(Earliest deadline First) algorithm executes the tasks based on their deadlines.That is, the task which has got the earliest deadline will get the highest priority to execute.This real time scheduling algorithm can be mapped with the fault tolerant scheduling(Primary Backup) by calling the subroutine function FT-EDF [2].

| Algorithm for Task Execution |
|---|
| 1. When any task is released on processor Pm, call the scheduler on Pm .<br>2. If portion 2 of task I is assigned on Pm then call scheduler on Pm-1 to execute portion 1 of task i.<br>3. Call the subroutine function FT-EDF.<br>4. Repeat for all the processors. |

Each processor will have its own scheduler and when the split task arrives, then the processor Pn+1 calls the processor Pn scheduler to execute the portion 1 of that task. The message transfer of scheduler or processor is not expensive as they share a common memory [3].Hence using EDF and Primary Backup model, all the tasks can be scheduled and executed in such a way that their deadlines are not missed.

## 5. MAPPING FT with EDF

Fault Tolerance is the ability to continue operating despite the failure of a limited subset of their hardware or software. There can be either hardware fault or software fault which may hinder the real time systems to meet their deadlines.

Fault detection can be achieved either through online or offline. One way of achieving is Primary-Backup model. Here, the tasks are assumed to be periodic and two instances of each task(a primary and a backup) are scheduled on a uni processor system. One of the restrictions to this approach is that the period of its preceding tasks. It also assumes that the execution time of the backup is shorter than that of the primary.

The following are the steps form the procedure used to implement the backup overloading algorithm

1. EDF Schedulability
   Check if all the tasks can be scheduled successfully using the earliest deadline first algorithm. If the schedulability test fails, then reject the set of tasks saying that they are not schedulable.

2. Searching for Timeslot
   When task Ti arrives, check each processor to find if the primary copy of the task can be scheduled between release time and deadline. Say it is scheduled on processor Pi.

3. Try Overloading
   Try to overload the backup copy on an existing backup slot on any processor other than Pi. The backups of 2 primary tasks that are scheduled on the same processor must not overlap. If the processor fails, it will not be possible to schedule the two backups simultaneously since they are on the same time slot (overloaded).

4. EDF algorithm
   If there is no existing backup slot that can be overloaded, then schedule the backup on the latest possible free slot depending upon the dead line of the task. The task with the earliest deadline is scheduled first.

5. De-allocation of backup
   If a schedule has been found for both the primary and backup copy for a task, commit the task, otherwise reject it. If the primary copy executes successfully, the corresponding backup copy is de-allocated.

6. Backup Execution.
   If there is a permanent or transient fault in the processor, the processor crashes and then all the backups of the tasks that were running on this system are executed on different processors.

## 5. CONCLUSION

The proposed algorithm SASA(Sort And Sequential Assignment) , is a real time task assignment algorithm for multi processor. It enhances the system performance by improving the time efficiency of the system. It further strives to achieve fault tolerance with task split. This approach can be implemented in any of the real time applications

## 6.REFERENCES